\documentclass[%
 reprint,
 amsmath,amssymb,
 aps,
]{revtex4-2}

\usepackage{graphicx}
\usepackage{dcolumn}
\usepackage{bm}

\usepackage{hyperref}
\hypersetup{hypertex=true,
			colorlinks=true,
			linkcolor=blue,
			anchorcolor=blue,
			citecolor=blue}

\begin{document}

\preprint{APS/123-QED}

\title{Multiphase curved boundary condition in lattice Boltzmann method}

\author{Yichen Yao}
\author{Yangsha Liu}
\author{Xingguo Zhong}
\author{Binghai Wen}
\email{oceanwen@gxnu.edu.cn}
\affiliation{%
Guangxi Key Lab of Multi-source Information Mining \& Security, Guangxi Normal University, Guilin 541004, China
}%
\affiliation{%
School of Computer Science and Engineering, Guangxi Normal University, Guilin 541004, China
}%

\begin{abstract}
The boundary treatment is fundamental for modeling fluid flows; especially, in the lattice Boltzmann method, the curved boundary conditions effectively improve the accuracy of single-phase simulations with complex-geometry boundaries. However, the conventional curved boundary conditions usually cause dramatic mass leakage or increase when they are directly used for multiphase flow simulations. We find that the principal reason is the absence of nonideal effect in the curved boundary conditions, followed by the calculation error. In this paper, incorporating the nonideal effect into the linear interpolation scheme and compensating for the interpolating error, we propose a multiphase curved boundary condition to treat the wetting boundaries with complex geometries. A series of static and dynamic multiphase simulations with large density ratio verify that the present scheme is accurate and ensures mass conservation. 
\end{abstract}

\maketitle

\section{Introduction}\label{sec1}
The lattice Boltzmann method (LBM) has already developed into an excellent numerical scheme for simulating complex fluid flows and phase transitions \cite{Chen1998,Li2016}. Since the boundary treatments play significant roles in the issues of accuracy, stability and efficiency, the lattice Boltzmann community has designed all kinds of boundary conditions \cite{Filippova1998,Mei1999,Bouzidi2001,Lallemand2003}. For a solid boundary, the first thing of a boundary condition (BC) in LBM is to calculate the missing distribution functions which are streaming from the boundary into the fluid. The halfway bounce-back boundary condition (HBBC) is popular in lattice Boltzmann simulations, because it is simple, robust, mass-conserving, and has formally second-order accuracy. However, HBBC has to locate the boundary in the middle of fluid-solid links; this degrades a curved boundary into a zigzag geometry, and then damages the accuracy of simulations inevitably  \cite{Ladd1994,Wen2015_Entorpy}. The requirement that captures the exact boundary geometry inspires the curved BCs, which usually aim for the second-order accuracy. For the boundaries with complex geometry, a few curved BCs have been proposed and widely used in single-phase lattice Boltzmann simulations. Filippova and Hanel \cite{Filippova1998} constructed a fictitious equilibrium distribution function of the solid node, and a linear interpolation was applied to calculate the distribution function bouncing back from the wall. Nevertheless, the scheme suffers from numerical instability when the relaxation time is close to 1 \cite{GuoZhaoliandShu2013}. Mei et al. \cite{Mei1999} eliminated the stability issue by replacing the velocity of the solid node with that of the secondary neighbor fluid node, when the boundary is closer to the fluid node. Bao et al. \cite{Bao2008} further improved the Mei scheme to satisfy mass conservation by redefining the density term of the wall nodes in the fictitious equilibrium distribution function. Guo et al. \cite{Guo2002} proposed a curved boundary treatment with high numerical stability by combining the non-equilibrium extrapolation scheme and spatial interpolation. From a mathematical point of view, Bouzidi et al. \cite{Bouzidi2001} proposed the interpolation boundary condition (IBC) by combining the bounce-back scheme and linear or quadratic interpolation. Lallemand and Luo \cite{Lallemand2003} extended the BCs to moving boundaries and verified the Galilean invariance. Yu et al. \cite{Yu2003} developed a unified version of the Bouzidi scheme, which maintains the geometric integrity of the curved wall and avoids the boundary treatment discontinuity of the previous scheme. Without interpolation or extrapolation, Tao et al. \cite{Tao2018} and Zhao et al. \cite{Zhao2019} realized single-node curved BC with the second-order accuracy. They promote the computational efficiency and are able to capture the real geometry in the simulations of porous media. The conventional curved boundary conditions effectively improve the accuracy of simulations of single-phase fluid flows with complex-geometry boundaries.

For multiphase flows, the boundary treatment is more complex, because various multiphase models are based on different theories, while the boundary has to reflect the surface wettability. Numerical simulation of multiphase flow is one of the most successful applications of LBM; especially, the sing-component two-phase model is very popular \cite{Li2016,Shan1993,Chen2014,Huang2015}. The pseudopotential model mimicked the long-range intermolecular interaction by a density-dependent interparticle potential. Theoretical analyses showed that its mechanical stability solution agrees with thermodynamics only if the effective mass takes the strict exponential function form \cite{Shan2006,Benzi2006}. Yuan and Schaefer \cite{Yuan2006} calculated the effective mass based on an EOS, and this enabled the pseudopotential model to incorporate the common EOSs. Huang et al. \cite{Huang2011} and Li et al. \cite{Li2012} investigated the performances of forcing terms in the pseudopotential model. Li et al. \cite{Li2013,Li2019} improved the stability and thermodynamic consistency of the pseudopotential model with a large density ratio and implemented a virtual-density contact angle scheme. Yang et al. \cite{Yang2020} analyzed the contact angle hysteresis at large Bond number. The nonideal effect can also be evaluated from the thermodynamic free energy \cite{Swift1995,Zhang2003}. Wen et al. \cite{Wen2017} directly evaluated the nonideal force by chemical potential and constructed a chemical-potential multiphase model, which meets thermodynamics and Galilean invariance. Then, a proportional coefficient was introduced to decouple the computational mesh from the momentum space and improved the model to simulate multiphase systems with very large density ratios and small spurious currents \cite{Wen2020}. A chemical-potential boundary condition was implemented to express the surface wettability, so that the contact angle can be linearly tuned by the chemical potential of the surface. During the multiphase simulation, the real-time contact angle can be accurately measured, and this enables the mechanical analyses at the three-phase contact line in real time \cite{He2020,Ji2021,Liu2022}. However, when the conventional curved BCs are directly used in multiphase lattice Boltzmann simulations, they cause dramatic mass leakage or increase, and usually collapse the evolutions, because they cannot deal with the large change of fluid density adjacent to a wetting boundary. Some studies directly compensated the leaking or increasing mass to the $0^{th}$ distribution function to enforce the mass conservation, and proposed the modified interpolation boundary condition (MIBC) \cite{Sanjeevi2018,Yu2020}. However, in some cases, the scheme causes the $0^th$ distribution function to be negative \cite{Yu2020}. Furthermore, it requires very large mass compensations and produces intense spurious currents (shown in sections 4.3 and 4.4), therefore the scheme is inaccurate and doesn't solve the underlying problems. Essentially, the mass problem in these curved BCs arises from the fact that the nonideal effect has not been considered properly.

In this paper, we incorporate the nonideal effect into the linear interpolation scheme and propose a multiphase curved BC. The simulating results demonstrate that this scheme not only guarantees the mass conservation, but also achieves high accuracy and small spurious currents. The paper is organized as follows. The second section describes the numerical methods including the lattice Boltzmann method, the chemical-potential multiphase model and the chemical-potential boundary condition. In the third section, we illustrate the present multiphase curved BC. The fourth section presents a series of verifications and test cases. Finally, the fifth section gives a summary of the paper.

\section{Numerical methods}\label{sec2}
\subsection{Lattice Boltzmann method}
Discretized fully in time, space and velocity, the lattice Boltzmann equation (LBE) with the multiple relaxation times can be concisely written as \cite{H1992,Lallemand2000}
\begin{equation}\label{eq1}
{f_i}\left( {\bm{x} + {{\bm{e}}_i}\delta t,{\rm{ }}t + \delta t} \right) - {f_i}\left( {\bm{x},{\rm{ }}t} \right) =  - {{\bf{M}}^{ - 1}} \cdot {\bf{S}} \cdot \left[ {{\bf{m}} - {{\bf{m}}^{{\rm{(eq)}}}}} \right],
\end{equation}
where ${f_i}(\bm{x},{\rm{ }}t)$ is the particle distribution function at time $t$ and lattice site $\bm{x}$, moving along the direction defined by the discrete velocity vector ${{\bm{e}}_i}$ with $i = 0,...,N$, $\bf{m}$ and $\bf{{m^{({\rm{eq}})}}}$ represent the velocity moments of the distribution functions and their equilibria, respectively; $\bf{M}$ is a matrix that linearly transforms between the distribution functions and the velocity moments, namely ${\bf{m}} = {\bf{M}} \cdot f$; and $f = {{\bf{M}}^{{\bf{ - 1}}}} \cdot {\bf{m}}$. For the two-dimensional nine-velocity (D2Q9) model on a square lattice (N = 8), the velocity moments are ${\bf{m}} = {{\rm{(}}\rho {\rm{, }}e{\rm{, }}\varepsilon {\rm{, }}{j_x}{\rm{, }}{q_x}{\rm{, }}{j_y}{\rm{, }}{q_y}{\rm{, }}{p_{xx}}{\rm{, }}{p_{xy}}{\rm{)}}^{\rm{T}}}$. The conserved moments are the density $\rho$ and the flow momentum ${\bm{j}} = {\rm{(}}{j_x}{\rm{,  }}{j_y}{\rm{)}} = \rho {\bm{u}}$, ${\bm{u}}$ is the local velocity. The equilibria of nonconserved moments depend only on the conserved moments: ${e^{(eq)}} =  - 2\rho  + \frac{3}{\rho }{\rm{(}}j_x^2 + {\rm{ }}j_y^2{\rm{)}}$, ${\varepsilon ^{(eq)}} = \rho  - \frac{3}{\rho }{\rm{(}}j_x^2 + {\rm{ }}j_y^2{\rm{)}}$, $q_x^{(eq)} =  - {j_x}$, $q_y^{(eq)} = {j_y}$, $p_{xx}^{(eq)} = \frac{1}{\rho }{\rm{(}}j_x^2 - {\rm{ }}j_y^2{\rm{)}}$, $p_{xy}^{(eq)} = \frac{1}{\rho }{\rm{(}}j_x^{}j_y^{}{\rm{)}}$. The transformation matrix is defined by 
\begin{equation}\label{eq2}
    {\bf{M}} = \left[ {\begin{array}{*{20}{r}}
1&1&1&1&1&1&1&1&1\\
{ - 4}&{ - 1}&{ - 1}&{ - 1}&{ - 1}&2&2&2&2\\
4&{ - 2}&{ - 2}&{ - 2}&{ - 2}&1&1&1&1\\
0&1&0&{ - 1}&0&1&{ - 1}&{ - 1}&1\\
0&{ - 2}&0&2&0&1&{ - 1}&{ - 1}&1\\
0&0&1&0&{ - 1}&1&1&{ - 1}&{ - 1}\\
0&0&{ - 2}&0&2&1&1&{ - 1}&{ - 1}\\
0&1&{ - 1}&1&{ - 1}&0&0&0&0\\
0&0&0&0&0&1&{ - 1}&1&{ - 1}
\end{array}} \right].
\end{equation}
The diagonal relaxation matrix has the nonnegative relaxation rates,  
\begin{equation}\label{eq3}
    {\bf{S}} = {\rm{diag(1, }}{s_e}{\rm{, }}{s_\varepsilon }{\rm{, 1, }}{s_q}{\rm{, 1, }}{s_q}{\rm{, }}{s_\nu }{\rm{, }}{s_\nu }{\rm{)}},
\end{equation}
in which ${s_e} = 1.64$, ${s_\varepsilon } = 1.54$, ${s_q} = 1.7$ \cite{McCracken2005}. The shear viscosity is $\nu  = \frac{1}{3}\left( {\frac{1}{{{s_v}}} - \frac{1}{2}} \right)$; in the present study, ${s_v} = 1/\tau$, and ${\tau} = 0.8$ unless otherwise specified.
The evolution of LBM includes two essential steps, namely collision and streaming; hence, the corresponding computations of LBE are performed as:
\begin{equation}\label{eq4}
{\rm{Collision:  }}{\tilde f_i}{\rm{(}}{\bm{x}},t{\rm{)}} - {f_i}{\rm{(}}{\bm{x}},t{\rm{)}} =  - {{\bf{{M}}}^{{\rm{ - 1}}}} \cdot {\bf{S}} \cdot \left[ {{\rm{ }}{\bf{m}} - {{\bf{m}}^{({\rm{eq}})}}} \right],
\end{equation}
\begin{equation} \label{eq5}
{\rm{Streaming:  \quad\quad}}{f_i}\left( {\bm{x} + {{\bm{e}}_i}\delta t,{\rm{ }}t + \delta t} \right) = {\tilde f_i}\left( {\bm{x},{\rm{ }}t} \right),\quad\quad
\end{equation}
where ${f_i}$ and ${\tilde f_i}$ denote pre-collision and post-collision states of the particle distribution functions, respectively. The dominant part of the computations, namely the collision step, is completely local, therefore the discrete equations are natural to parallelize.

\subsection{Chemical-potential multiphase model}
For a nonideal fluid system, following the classical capillarity theory of van der Waals, the free energy functional within a gradient-squared approximation is written as \cite{Swift1995,Rowlinson1982,Swift1996,Wen2015_EL,Jamet2002}
\begin{equation}\label{eq6}
    \Psi =\int{\left[ \psi (\rho )+\frac{\kappa }{2}|\nabla \rho {{|}^{2}} \right]}d\mathbf{x},
\end{equation}
where the first term represents the bulk free-energy density and the second term describes the contribution from density gradients in an inhomogeneous system, and $\kappa $  is the surface tension coefficient. The general equation of state and chemical potential can be defined by the free energy density,
\begin{equation}\label{eq7}
    {{p}_{0}}=\rho {\psi }'(\rho )-\psi (\rho ),
\end{equation}
and
\begin{equation}\label{eq8}
    \mu ={\psi }'(\rho )-\kappa {{\nabla }^{2}}\rho,
\end{equation}
respectively. The chemical potential is the partial molar Gibbs free energy at constant pressure \cite{Jamet2002}. The gradient of chemical potential is the driving force for isothermal mass transport. The movement of molecules from higher to lower chemical potential is accompanied by a release of free energy, and the chemical or phase equilibrium is achieved at the minimum free energy. With respect to the ideal gas pressure, the nonideal force can be evaluated by a chemical potential \cite{Wen2017}
\begin{equation}\label{eq9}
    \bm{F}=-\rho \nabla \mu +c_{s}^{2}\nabla \rho. 
\end{equation}

Solving the linear ordinary differential equations \eqref{eq7} and \eqref{eq8}, and substituting a specific EOS, the chemical potentials of some widely used EOSs can be obtained analytically. The present study applies the famous Peng-Robinson (PR) EOS 
\begin{equation}\label{eq10}
    {{p}_{0}}=\frac{\rho RT}{1-b\rho }-\frac{a\alpha (T){{\rho }^{2}}}{1+2b\rho -{{b}^{2}}{{\rho }^{2}}},
\end{equation}
and its chemical potential is
\begin{equation}\label{eq11}
\begin{aligned}
\mu _{{}}^{\text{PR}}&=RT\ln \frac{\rho }{1-b\rho }-\frac{a\alpha (T)}{2\sqrt{2}b}\ln \frac{\sqrt{2}-1+b\rho }{\sqrt{2}+1-b\rho }+\frac{RT}{1-b\rho }\\
&-\frac{a\alpha (T)\rho }{1+2b\rho -{{b}^{2}}{{\rho }^{2}}}-\kappa {{\nabla }^{2}}\rho,
\end{aligned}
\end{equation}
where $R$ is the gas constant, $a$ is the attraction parameter, $b$ is the volume correction parameter, and the temperature function is $\alpha (T) = {\left[ {1 + \left( {0.37464 + 1.54226\omega  - 0.26992{\omega ^2}} \right)\left( {1 - \sqrt {T/{T_c}} } \right)} \right]^2}$. In the present simulations, the parameters are given by $a = 2/49$,  $b = 2/21$, and $R = 1$. The acentric factor $\omega$ is 0.344 for water. To make the numerical results closer to the actual physical properties, we define the reduced variables ${T_r} = T/{T_c}$  and ${\rho _r} = \rho /{\rho _c}$  , in which  ${T_c}$ is the critical temperature and  ${\rho _c}$ is the critical density.

A proportional coefficient $k$ is introduced to decouple the computational mesh from the momentum space \cite{Wen2020} and relates the length units of the two space, $\delta \hat x = k\delta x$ . Here the quantities in the mesh space are marked by a superscript. Following the dimensional analysis, the chemical potential in the mesh space can be evaluated by \cite{Wen2017,Wen2020}  
\begin{equation}\label{eq12}
    \hat{\mu }={{k}^{2}}{\psi }'(\rho )-\hat{\kappa }{{\hat{\nabla }}^{2}}\rho.
\end{equation} 
This approach greatly improves the stability of the chemical-potential multiphase model, and the transformation holds the mathematical equivalence and has no loss of accuracy. 

The resulted external force is incorporated into LBE by a forcing technique. Here, the exact difference method is used and the body force term  is simply equal to the difference of equilibrium distribution functions before and after the nonideal force acting on the fluid during a time step \cite{Kupershtokh2009}:
\begin{equation}\label{eq13}
{{F}_i} = f_i^{(eq)}(\rho ,\bm{u} + \delta \bm{u}) - f_i^{(eq)}(\rho ,\bm{u}),
\end{equation}
where ${\bm{u}}$ is the fluid velocity, $\delta{\bm{u}} = \delta t\bm{F}/\rho $. and $f_i^{(eq)}$ is the equilibrium distribution function,
\begin{equation}\label{eq14}
f_i^{(eq)}({\bm{x}},t) = {\omega _i}\rho ({\bm{x}},t)\left[ {1 + \frac{{({{\bm{e}}_i} \cdot {\bm{u}})}}{{c_s^2}} + \frac{{{{({{\bm{e}}_i} \cdot {\bm{u}})}^2}}}{{2c_s^4}} - \frac{{{{({\bm{u}})}^2}}}{{2c_s^2}}}, \right]
\end{equation}
where the sound speed is $c_s^{} = {{\sqrt 3 } \mathord{\left/
 {\vphantom {{\sqrt 3 } 3}} \right.
 \kern-\nulldelimiterspace} 3}$. Accordingly, the macroscopic fluid velocity is redefined as: ${\bm{v}} = {\bm{u}} + \delta t{\bm{F}}/(2\rho )$.

\subsection{Chemical-potential boundary condition}\label{sec2.3}
The chemical potential can effectively indicate the wettability of a solid surface [26]. Since the present multiphase model is driven by a chemical potential, the implementation of the chemical potential boundary condition is simple and quite natural. To specify the wettability of a solid surface, one can assign a specific chemical potential to the solid nodes of the surface. The specific chemical potential gives rise to a chemical-potential gradient between the solid surface and the neighbor fluid, reflecting the nonideal interaction between them and resulting in the wetting phenomena. In order to calculate the density gradient near the boundary, the boundary condition needs to estimate the densities on the solid nodes adjacent to the fluid, which include two or three layers of solid nodes when the gradient is calculated by the central difference methods with the fourth or sixth order accuracy, respectively. The densities on these layers of solid nodes can be calculated based on the nearest neighbor nodes \cite{Liu2022},
\begin{equation}\label{eq15}
\rho \left( {{{\bm{x}}_s}} \right) = \frac{{\sum\limits_i {{\omega _i}} \rho \left( {{{\bm{x}}_s} + {{\bm{e}}_i}{\delta _t}} \right){s_w}}}{{\sum\limits_i {{\omega _i}} {s_w}}}
\end{equation}
where ${\bm{x_s}} + {{\bm{e}}_i}{\delta _t}$ indicates the adjoining nodes, and ${s_w}$ is a switching function. For the first layer of solid nodes, ${s_w} = 1$ when ${\bm{x_s}} + {{\bm{e}}_i}{\delta _t}$ is a fluid node; for the second or third layer of nodes, ${s_w} = 1$ when ${\bm{x_s}} + {{\bm{e}}_i}{\delta _t}$ is in the first or second layer, respectively; otherwise, ${s_w} = 0$. When the chemical-potential boundary condition is implemented on a horizontal substrate, the above equation can be simplified to a simple weighted average scheme based on the densities of the neighbor layer \cite{Wen2017}.

\section{Curved boundary condition for multiphase flow}\label{sec3}
An important function of a boundary condition in LBM is to evaluate the missing distribution functions, which, in concept, stream from the boundary to the fluid. For single-phase flows, fluid density can be regarded as uniform, nevertheless the local density would fluctuate slightly due to coupling to pressure. Without phase transitions, the collisions of distribution functions do not involve nonideal effects. Thus, the conventional curved boundary conditions, especially those based on interpolation algorithms, work very well, such as linear interpolation, quadratic interpolation, fictitious distribution function, multiple reflection boundary conditions and so on \cite{Filippova1998,Bouzidi2001,Ginzburg2003}. The multiphase lattice Boltzmann simulations are quite different. Because the Boltzmann equation assumes that the particles are uncorrelated prior to the collisions, LBE cannot directly describe phase transitions. A multiphase model has to evaluate the nonideal force and add it into the LBE collision term as an external force. A boundary condition can produce new distribution functions, so it acts as a collision step. The nonideal effect should be considered in order to reflect the surface wettability; otherwise, the errors are intolerable. Furthermore, in the typical water/vapor system, the liquid-gas density ratio can reach around hundreds to thousands of times, and the density profile across the phase transition region is highly nonlinear; consequently, the boundary conditions result in sizable calculating errors at three-phase contact regions. Even at the liquid-solid and gas-solid interfaces, the wetting boundary would remarkably change the adjacent fluid density and leads to interpolating errors. A distribution function can be viewed as a mass component in LBM \cite{Wen2014}, these errors cause abnormal mass leakage or increase and then crash the evolution in the multiphase simulations (please refer to Fig. \ref{FIG3} to see a typical example). Therefore, the nonideal effect and the interpolating error must be handled properly before a curved boundary condition is implemented in multiphase simulations.
\subsection{Nonideal effect in curved boundary condition}\label{sec3.1}
\begin{figure}[h]
\includegraphics[width=8.4cm]{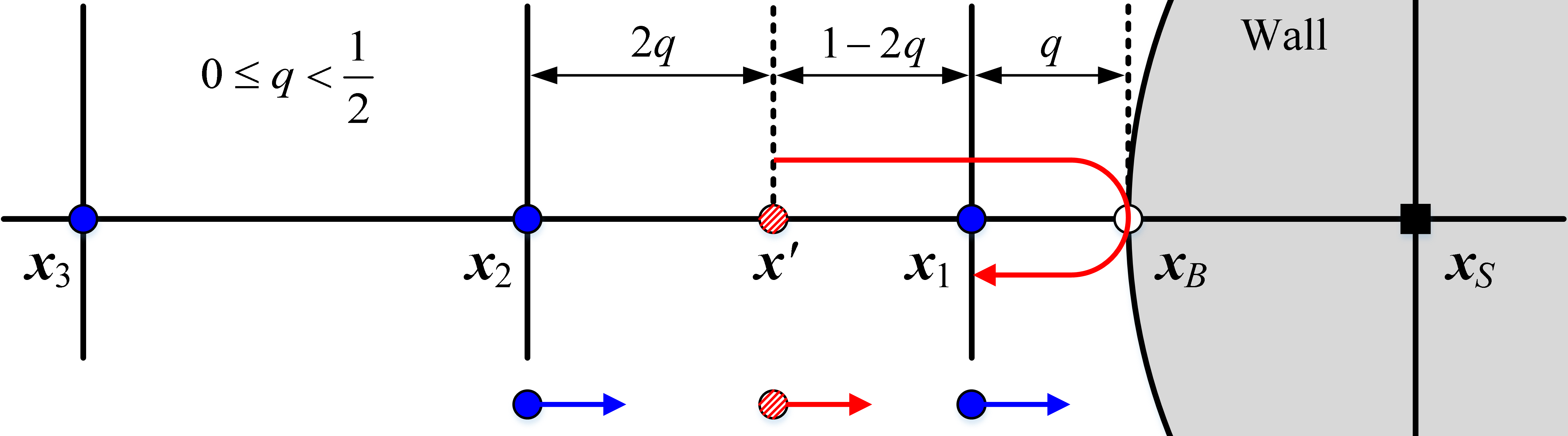}
\caption{\label{FIG1} Schematic diagram of the multiphase curved boundary condition for $ 0 \le q < \frac{1}{2}$.}
\end{figure}

\begin{figure}[h]
	\centering
		\includegraphics[width=8.6cm]{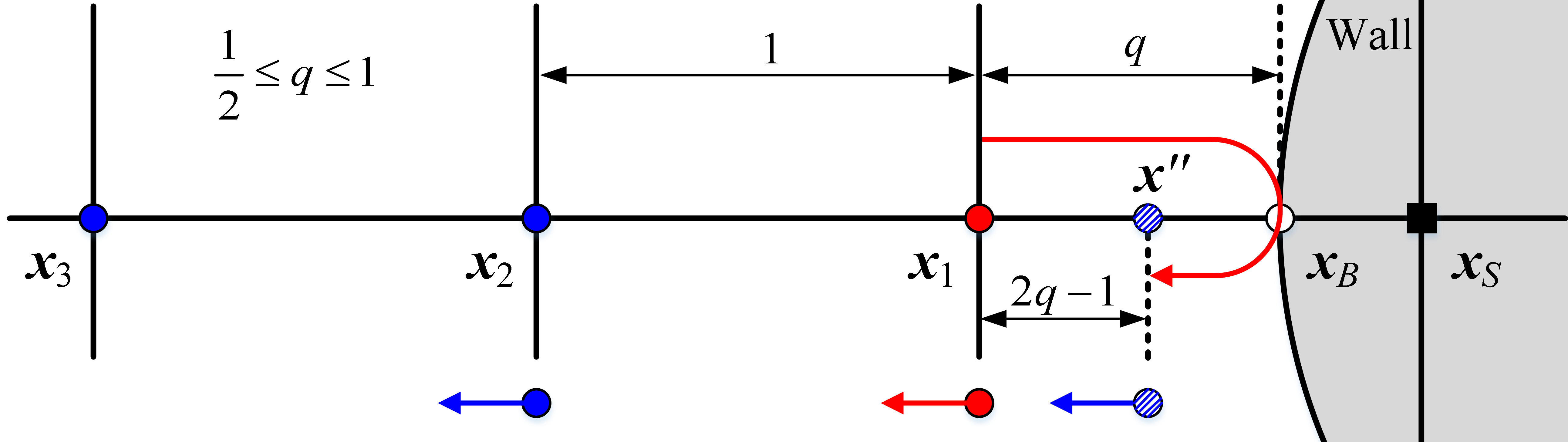}
	\caption{Schematic diagram of the multiphase curved boundary condition for $\frac{1}{2} \le q \le 1$. }
	\label{FIG2}
\end{figure}

The present scheme is based on the interpolation boundary condition, which combines the linear interpolation and the bounce-back scheme \cite{Bouzidi2001}. Figs. \ref{FIG1} and \ref{FIG2} are the schematic diagrams of the present multiphase curved BC. A solid boundary is locating between a solid node ${\bm{x}_s}$ and a boundary-fluid node ${\bm{x}_1}$, which is a fluid node linked directly to the boundary.  If $i$ denotes  the discrete direction from the node ${\bm{x}_1}$ to the node ${\bm{x}_s}$, then the task of the boundary condition is to calculate the distribution function ${f_{\bar i}}({{\bm{x}}_{\rm{1}}},t + 1)$ for the next time step. A multiphase curved BC needs to correctly treat the nonideal interactions between the wetting boundary and its adjacent liquid.

In the schematic diagram of Fig. \ref{FIG1}, the distance from the boundary to the boundary-fluid node is $ 0 \le q < \frac{1}{2}$, and the fictitious node ${\bm{x}'}$ is $1 - 2q$ lattice unit away from ${\bm{x}_1}$. If the fictitious distribution function ${\tilde f_i}({\bm{x}'},t)$ at the time step $t$ is known, the particle population streams towards ${\bm{x}_s}$ passing ${\bm{x}_1}$, collides with the boundary and bounces back, and then travels to ${\bm{x}_1}$ in a single time step; finally, it becomes ${f_{\bar i}}({{\bm{x}}_{\rm{1}}},t + 1)$. The fictitious distribution function can be calculated directly by the linear interpolation of the distribution functions ${\tilde f_i}({{\bm{x}}_{\rm{1}}},t)$ and ${\tilde f_i}({{\bm{x}}_{\rm{2}}},t)$ in a single-phase environment \cite{Bouzidi2001}. However, in multiphase simulations, the boundary condition has to consider the nonideal effect that a distribution function collides with the wetting boundary. This nonideal effect can be embodied by the nonideal force ${\bm{F}}({{\bm{x}}_{\rm{1}}},t)$ on the boundary-fluid node ${\bm{x}_1}$. Let’s observe the two distribution functions that constitute the interpolation. Obviously, during the time step $t$, ${\tilde f_i}({{\bm{x}}_{\rm{1}}},t)$ has already collided with the boundary and contained the nonideal effect. But ${\tilde f_i}({{\bm{x}}_{\rm{2}}},t)$ cannot reach the boundary in the time step $t$, so that it neither collides with the boundary nor contains the nonideal effect. The part of nonideal effect should be complemented according to the proportion of ${\tilde f_i}({{\bm{x}}_2},t)$ in the interpolation. Thus, we can design the multiphase curved boundary condition for $ 0 \le q < \frac{1}{2}$ by integrating the linear interpolation, the bounce-back scheme and the nonideal effect,
\begin{equation}\label{eq16}
\begin{aligned}
  {f_{\bar i}}({{\bm{x}}_{\rm{1}}},{\rm{ }}t + 1) &= {\rm{2}}q{\kern 1pt} {\tilde f_i}({{\bm{x}}_1},{\rm{ }}t) + (1 - 2q){\tilde f_i}({{\bm{x}}_{\rm{2}}},{\rm{ }}t) \\
&+ (1 - 2q){{F}_i}({{\bm{x}}_1},{\rm{ }}t).
\end{aligned}
\end{equation}

In the schematic diagram of Fig. \ref{FIG2}, the distance from the boundary to the boundary-fluid node is $\frac{1}{2} \le q \le 1$. If the above scheme is still used, the fictitious node would locate beyond ${\bm{x}_1}$, and the calculation will degenerate into an extrapolation. As a more accurate scheme, the interpolation can be established on the distribution functions of the $t + 1$ time step. After the local collision, ${\tilde f_i}({{\bm{x}}_{\rm{1}}},t)$ streams towards ${\bm{x}_s}$, collides with the boundary and bounces back, and then travels to the fictitious node ${\bm{x}''}$ in a single time step, which is $2q - 1$ lattice unit away from ${\bm{x}_1}$; finally, it becomes the fictitious distribution function ${f_{\bar i}}({\bm{x}''},t + 1)$. ${f_{\bar i}}({{\bm{x}}_{\rm{1}}},t + 1)$ is calculated by the interpolation of ${f_{\bar i}}({{\bm{x}}_2},t + 1)$ and ${f_{\bar i}}({\bm{x}''},t + 1)$, and the nonideal effect that a distribution function collides with the wetting boundary has also to be taken into consideration. Obviously, different from ${f_{\bar i}}({\bm{x}''},t + 1)$, ${f_{\bar i}}({{\bm{x}}_2},t + 1)$ has neither collided with the boundary in the time step $t$ nor contained the nonideal effect of the wetting boundary. The part of nonideal force should also be complemented according to the proportion of ${f_{\bar i}}({{\bm{x}}_2},t + 1)$ in the interpolation. To avoid referring to the fictitious node, we use the post-collision distribution functions at the $t$ time step; thus, the multiphase curved boundary condition for $\frac{1}{2} \le q \le 1$ can be written as
\begin{equation}\label{eq17}
\begin{aligned}
 {f_{\bar i}}({{\bm{x}}_{\rm{1}}},{\rm{ }}t + 1) &= \frac{{\rm{1}}}{{{\rm{2}}q}}{\tilde f_i}({{\bm{x}}_1},{\rm{ }}t) + \frac{{{\rm{(}}2q - {\rm{1)}}}}{{{\rm{2}}q}}{\kern 1pt} {\tilde f_{\bar i}}({{\bm{x}}_1},{\rm{ }}t) \\
 &+ \frac{{{\rm{(}}2q - {\rm{1)}}}}{{{\rm{2}}q}}{{F}_i}({{\bm{x}}_1},{\rm{ }}t).
 \end{aligned}
\end{equation}
\subsection{Mass modification for interpolating error}\label{sec3.2}
Although complementing the nonideal effect has greatly reduced the error, it is inevitable that the interpolation itself causes a certain degree of errors when the density profile is nonlinear. The interpolating errors are noticeably fluctuant at the beginning of a multiphase simulation and become very small after the initial stage. They will influence the chemical potential and then disturb the phase transition and equilibrium. The mass fluctuation caused by the interpolating error is compensated as the local and static mass component in every time step \cite{Sanjeevi2018,Yu2020}, 
\begin{equation}\label{eq18}
{f_0}({\bm{x}_{1}},{\rm{ }}t + 1) = {f_0}({\bm{x}_1},{\rm{ }}t + 1) + \sum {\left( {{{\tilde f}_i}({\bm{x}_1},{\rm{ }}t) - {f_{\bar i}}({\bm{x}_{\rm{1}}},{\rm{ }}t + 1)} \right)},
\end{equation} 
where the sum includes all of the differences between the outflow and inflow distribution functions on the fluid-solid links of each boundary-fluid nodes. This modification guarantees the mass conservation without violating the conservation of momentum.

The previous studies used MIBC in multiphase simulations with curved wetting boundaries. After the interpolation of IBC, MIBC uses Eq. \eqref{eq18} alone without the consideration of nonideal effect \cite{Sanjeevi2018,Yu2020}. Thus, the mass compensations in every time step are the total errors of the nonideal effect and the interpolation, which are very large as shown in Figs. \ref{FIG7}(a) and \ref{FIG8}(a). Whereas, since the nonideal effect is properly treated, the mass modifications in the present scheme only involve the interpolating errors in every time step. After the initial stage, they gradually reduce to approach 0, as shown in Figs. \ref{FIG7}(b) and \ref{FIG8}(b).

When the boundary locates exactly in the middle of a fluid node and a solid node, namely $q = \frac{1}{2}$, the last two terms in Eq. \eqref{eq17}, which originate from the fictitious distribution function and the nonideal effect, are equal to zero. Thus, Eq. \eqref{eq17} becomes ${f_{\bar i}}({{\bm{x}}_{\rm{1}}},{\rm{ }}t + 1) = {\tilde f_i}({{\bm{x}}_1},{\rm{ }}t)$, and Eq. \eqref{eq18} can be canceled. In other words, IBC, MIBC and the present scheme are identical to HBBC when the distance $q$ is equal to 0.5 exactly. Since it only refers to ${\tilde f_i}({{\bm{x}}_1},{\rm{ }}t)$ without interpolation, which collides with the boundary and bounces back in the time step $t$, HBBC has already considered the nonideal effect and involved no interpolating error. However, when a simulation includes some complex-geometry boundaries, HBBC replaces the real curved boundary by a zigzag geometry, which damages the computational accuracy definitely \cite{Ladd1994,Wen2015_Entorpy}.

\section{Numerical simulations and discussion}\label{sec4}
In this section, a series of simulations are performed to validate the effectiveness of the present multiphase curved BC. The sections \ref{sec4.1} and \ref{sec4.2} verify the mass conservation of the present scheme by simulating a drop on a flat or a curved wetting substrate. The section \ref{sec4.3} quantitatively analyzes the nonideal effect and the interpolating error. The section \ref{sec4.4} compares the spurious currents of the present scheme and MIBC. The simulations in these sections apply the rectangular computation domains, whose width and height are 500 and 300 lattice units for the flat substrates and 500 and 600 lattice units for the curved substrates, respectively. The section \ref{sec4.5} simulates drops falling on a curved wetting substrate and compares the results with the benchmarks from the experiments and simulations \citep{Bakshi2007,Fakhari2017}. The width and height of the computational domain are 600 and 1200 lattice units, respectively. The density fields are initialized by using the equation \cite{Li2016,Huang2011}
\begin{figure*}
	\centering
		\includegraphics[width=14.1cm]{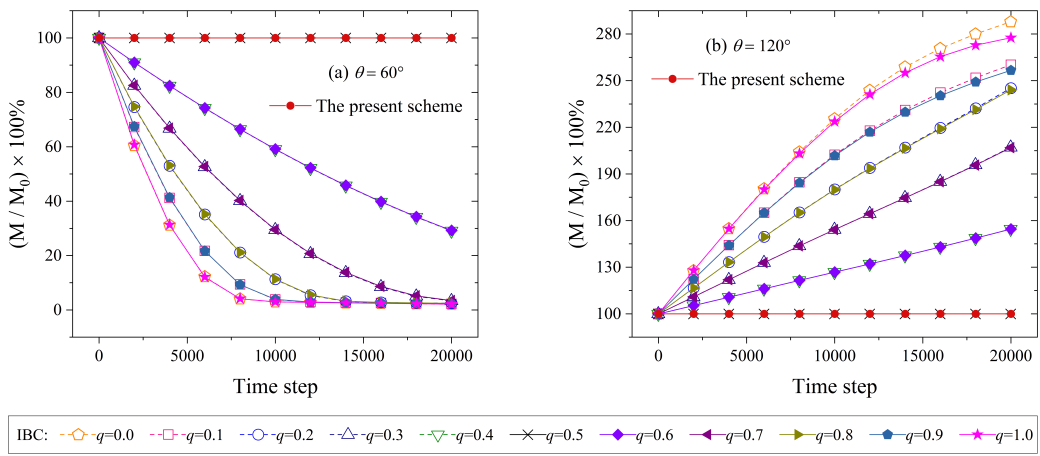}
	\caption{The mass change of drops on (a) a hydrophilic substrate with $\theta = 60^\circ$ and (b) a hydrophobic substrate with $\theta = 120^\circ$.}
	\label{FIG3}
\end{figure*}
\begin{figure*}
	\centering
		\includegraphics[width=14cm]{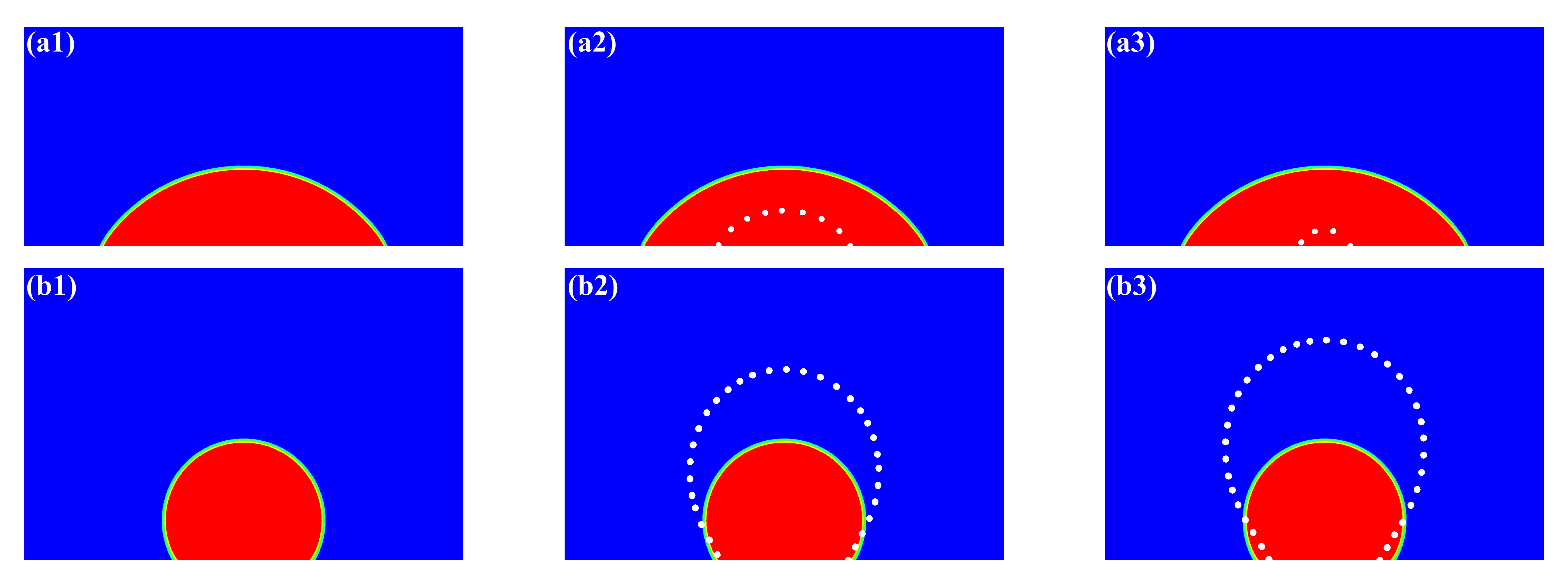}
	\caption{The dynamic evolutions of a drop on (a) a hydrophilic substrate with $\theta = 60^\circ$ and $q$ = 0.3, and (b) a hydrophobic substrate with $\theta = 120^\circ$ and $q$ = 0.8. The drop contours are simulated by using the present scheme, whereas the dotted lines are simulated by using IBC. The evolution times are 0, 10000 and 15000 time steps for (a1) $\sim$ (a3) and (b1) $\sim$ (b3), respectively.}
	\label{FIG4}
\end{figure*}
\begin{equation}\label{eq19}
\rho \left( {x,y} \right) = \frac{{{\rho _g} + {\rho _l}}}{2} + \frac{{{\rho _g} - {\rho _l}}}{2}\tanh \left[ {\frac{{2\left( {r - {r_0}} \right)}}{W}} \right],    
\end{equation}
where ${\rho _g}$  and ${\rho _l}$ are the two-phase coexistence densities calculated by the Maxwell equal-area construction, ${W = 10}$ is the initial interface width,  ${r_0}$ is the initial radius the droplet and $r = \sqrt {{{(x - {x_0})}^2} + {{(y - {y_0})}^2}} $ . The curvature radius of the curved substrate is 100 lattice units. The drop radii take 50, 70 and 100 lattice units for the flat substrate, the curved substrate and the drop falling cases, respectively. The periodic boundary condition is applied at the left and right boundaries, and HBBC is applied at the upper boundary. IBC, MIBC and the present scheme are implemented on the substrates, and the chemical-potential BC is used to realize the surface wettability. The contact angles of the substrates are assigned 60$^\circ$ for hydrophilic and 120$^\circ$ for hydrophobic substrates, besides another 150$^\circ$ hydrophobic substrate for the drop falling.

\subsection{Verification on flat wetting substrates}\label{sec4.1}
A sessile drop sitting on a flat wetting substrate is simulated to verify the mass conservation. The distances from the substrate to the horizontal lattice line take the successive values from 0 to 1 with a step length of 0.1. $\rm{M_0}$ and $\rm{M}$ indicate the initial and dynamic masses of the whole flow field, respectively. Fig. \ref{FIG3} presents clearly that, for all values of the distance $q$, the masses of flow field with the present BC are always constant and are identical to the results by HBBC (namely $q$ = 0.5), no matter on the hydrophilic or hydrophobic substrate. As a comparison, IBC is also used in the simulations. On the hydrophilic substrate, IBC causes dramatic mass leakage for all values of the distance $q$ except 0.5 as shown in Fig. \ref{FIG3}(a). Noticeably, the farther from $q$ = 0.5, the more serious the mass leakage, and they are almost symmetrical relative to $q$ = 0.5. At the time step 20000, the residual masses for $q$ = 0, 0.1, 0.2, 0.8, 0.9 and 1.0 have already reduced to about 2$\%$ of the initial mass, which indicates that all the liquid has already evaporated and the flow field turns into gas phase wholly. On the hydrophobic substrate, IBC produces a dramatic mass increase for all $q$ values except 0.5 as shown in Fig. \ref{FIG3}(b). Similarly, the farther from $q$ = 0.5, the more the mass increase, and they are almost symmetrical relative to $q$ = 0.5. At the time step 20000, the soaring masses for $q$ = 0.0, 0.1, 0.2, 0.8, 0.9 and 1.0 have already exceeded 240$\%$ of the initial mass, thus the whole flow field is turning into liquid phase gradually.
\begin{figure*}
	\centering
		\includegraphics[width=16.5cm]{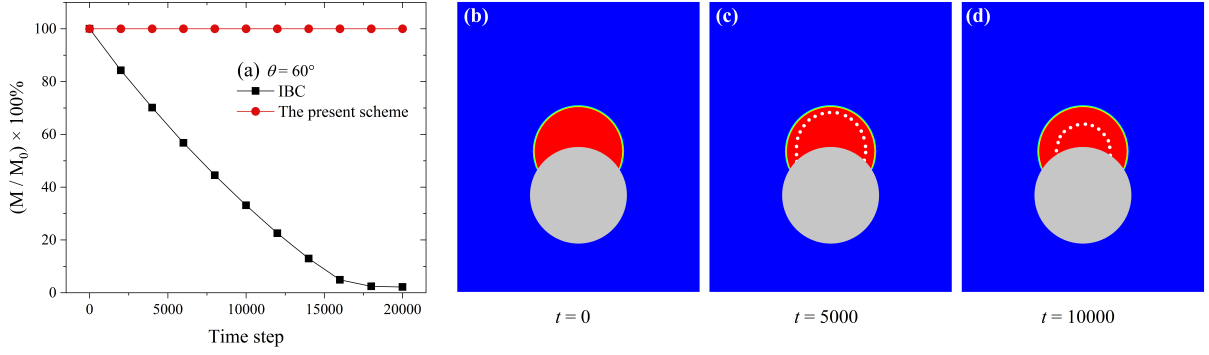}
	\caption{(a) The mass change of a drop on a hydrophilic curved substrate with $\theta = 60^\circ$. The evolving drops simulated by the present scheme (contours) and IBC (dotted lines) are compared at the time step (a) 0, (b) 5000 and (c) 10000.}
	\label{FIG5}
\end{figure*}
\begin{figure*}
	\centering
		\includegraphics[width=16.5cm]{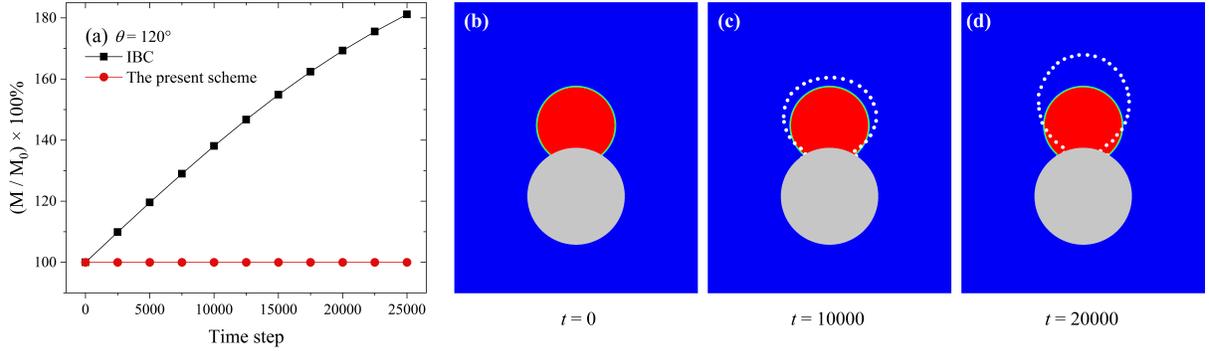}
	\caption{(a) The mass change of a drop on a hydrophobic curved substrate with $\theta = 120^\circ$. The evolving drops simulated by the present scheme (contours) and IBC (dotted lines) are compared at the time step (a) 0, (b) 10000 and (c) 20000.}
	\label{FIG6}
\end{figure*}

The changes of the drop sizes during the evolutions are illustrated in Fig. \ref{FIG4}, in which the subfigures (a1) $\sim$ (a3) apply the hydrophilic substrate with $\theta = 60^\circ$ and the subfigures (b1) $\sim$ (b3) apply the hydrophobic substrate with $\theta = 120^\circ$. During the evolutions of the multiphase system, the contour of the drop keeps the same when the present BC is applied. The dotted lines represent the results simulated by IBC. It is clear that IBC notably makes the drop shrink on the hydrophilic substrate and grow on the hydrophobic substrate.

The above simulations demonstrate that IBC, which is competent for treating complex boundaries in the simulations of single-phase flows, cannot be directly applied in multiphase flow environments. After considering the nonideal effect and compensating for the interpolating error, the present scheme meets the mass conservation on both hydrophilic and hydrophobic flat substrates, no matter the distances between the substrate and the horizontal lattice line.

\subsection{Verification on curved wetting substrates}\label{sec4.2}
Different from the simulations in the section \ref{sec4.1}, in which the flat substrate has a constant distance $q$ away from the horizontal lattice line, this section verifies the mass conservation of a sessile drop sitting on a curved wetting substrate. Since the curved boundary can involve all kinds of the distance $q \in [0,1]$ , the simulations in this section reflect the overall effect of a multiphase curved BC.
\begin{figure*}
	\centering
		\includegraphics[width=13.8cm]{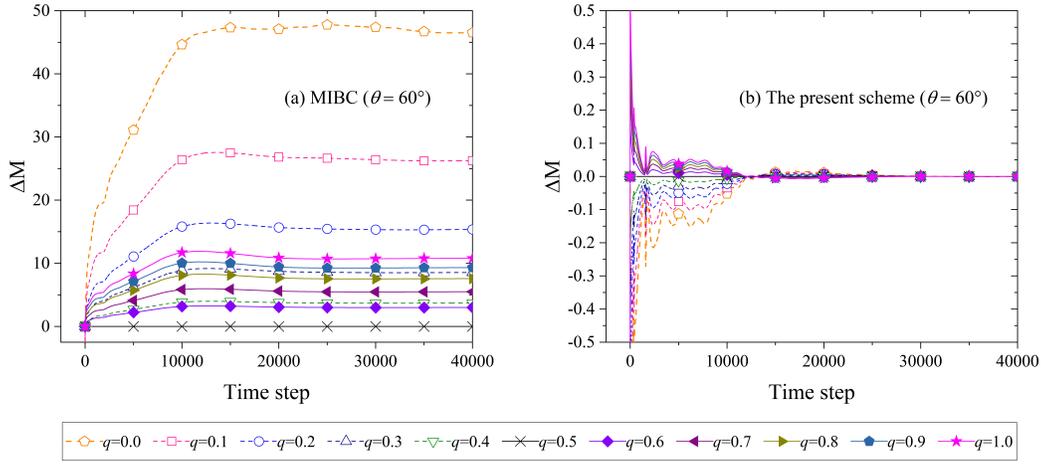}
	\caption{The compensating mass during the simulations of a drop on a hydrophilic substrate with $\theta $ = 60$^\circ$ by (a) MIBC and (b) the present scheme.}
	\label{FIG7}
\end{figure*}
\begin{figure*}
	\centering
		\includegraphics[width=13.8cm]{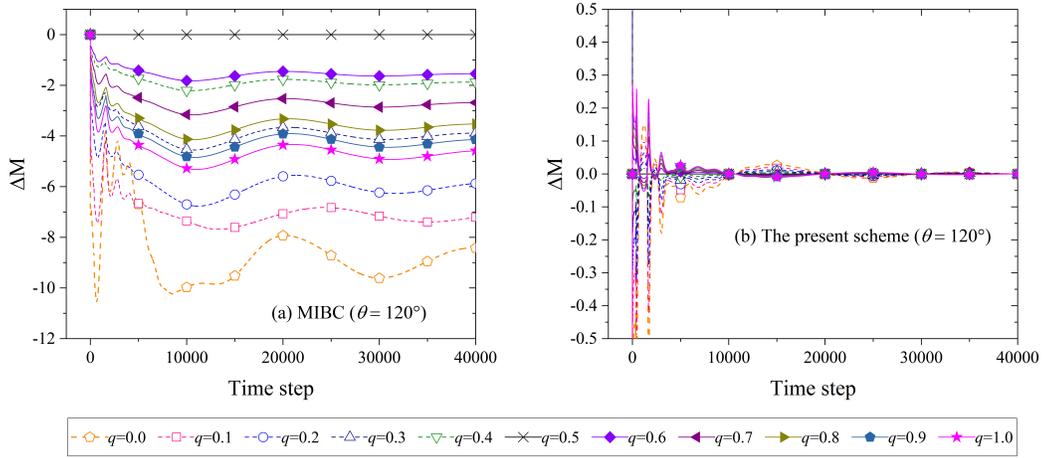}
	\caption{The compensating mass during the simulations of a drop on a hydrophobic substrate with $\theta $ = 120$^\circ$ implemented by (a) MIBC and (b) the present scheme.}
	\label{FIG8}
\end{figure*}

Fig. \ref{FIG5}(a) presents the mass changes of the flow field during the dynamic evolutions of a drop on a hydrophilic curved substrate with $\theta = 60^\circ$. The simulation with the present scheme keeps the mass constant. Whereas, IBC makes the mass reduce dramatically, and at the time step 20000, the residual mass has already reduced to about 2$\%$ of the initial mass, which suggests that the whole flow field has already turned into gas phase. he subfigures (b) $\sim$ (d) compare the drop size during the dynamical evolutions, and the contours and the dotted lines represent the simulating results by using the present scheme and IBC, respectively. Obviously, the drop in the flow field with IBC shrinks to fade. Fig. \ref{FIG6}(a) presents the mass changes of the flow field during the dynamic evolutions of a drop on a hydrophobic curved substrate with $\theta = 120^\circ$. The simulation with the present scheme keeps the mass constant, but that by using IBC makes the mass increase notably.  At the time step 25000, the total mass has already increased to more than 180$\%$ of the initial mass, and the whole flow field is turning into liquid phase. The subfigures (b) $\sim$ (d) display the visual comparisons, in which the contours and the dotted lines represent the drops simulated by the present scheme and IBC, respectively. Remarkably, the drop in the flow field with IBC grows quickly. These simulations confirm that the present scheme satisfies mass conservation, and IBC cannot be applied directly in a multiphase environment.

\subsection{Quantification of nonideal effect and interpolating errors}\label{sec4.3}
MIBC combines IBC and the Eq. \eqref{eq18}, and is equivalent to artificially compensate for all of the nonideal effect and the interpolating errors. Whereas, the present scheme takes into account the nonideal effect through Eqs. \eqref{eq16} and \eqref{eq17}, and only compensates for the interpolating error. In order to distinguish the influences caused by the nonideal effect and the interpolating error, the quantitative analyses are conducted by calculating the compensating mass $\Delta$M in every time step in the simulations with MIBC and the present scheme. 

Figs. \ref{FIG7}(a) and \ref{FIG8}(a) present the compensating masses of MIBC in the simulations of a drop on the wetting substrates, which include both the nonideal effect and the interpolating error. It is obvious that the absolute values of the artificial compensation for MIBC are very larger, and they are even close to -10 and 50 in a single time step for the hydrophobic and hydrophilic substrates, respectively. On the contrary, Figs. \ref{FIG7}(b) and \ref{FIG8}(b) show clearly that the compensating masses required by the present scheme, which only involves the interpolating error, are far less than 1. They appear mainly in the initial phase, fluctuate up and down, and converge to almost zero quickly. Comparing the results of MIBC and the present scheme manifests that the errors from the nonideal effect are much more than that from the interpolation, and the considerations of the nonideal effect in Eqs. \eqref{eq16} and \eqref{eq17} are accurate and stable. In other words, the present scheme minimizes the artificial intervention in the evolution of the flow field.

\subsection{Spurious currents of drops on flat or curved substrates}\label{sec4.4}
The spurious currents are a nonphysical phenomenon in lattice Boltzmann simulation for multiphase flow, and the boundary treatment plays an important role in spurious currents \cite{Shan2006,Wen2017}. This section compares the spurious currents caused by MIBC and the present scheme.
\begin{figure}[h]
	\centering
		\includegraphics[width=7cm]{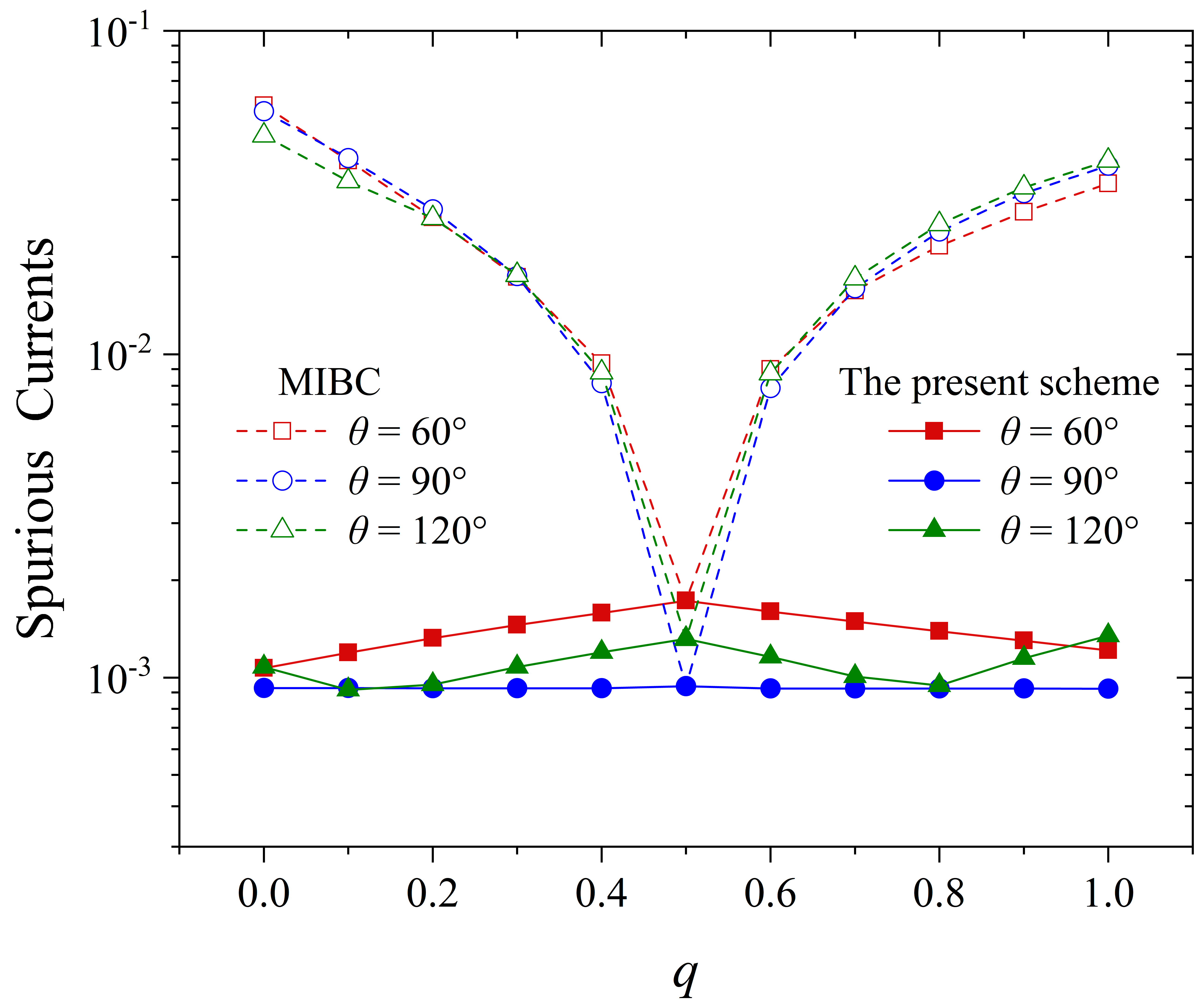}
	\caption{The spurious currents of a drop on a flat wetting substrate with the contact angles $\theta = 60^\circ$, $90^\circ$ and $120^\circ$.}
	\label{FIG9}
\end{figure}

The simulations of a drop on various flat wetting substrates are performed, and the contact angle $\theta = 60^\circ$, $90^\circ$ and $120^\circ$ represent the hydrophilic, neutral and hydrophobic substrates, respectively. Fig. \ref{FIG9} shows that the spurious currents from MIBC are sizable except $q$ = 0.5. They approach ${10^{ - 2}}$ for $q$ = 0.4 and 0.6, and increase even to $5 \times {10^{ - 2}}$ when the distances are more away from 0.5. Strikingly, the spurious currents from the present scheme are about one order magnitude lower than those from MIBC except $q$ = 0.5. Even more amazing is that most of the spurious currents from the present scheme are less than those from HBBC. 

\begin{figure}[h]
	\centering
		\includegraphics[width=7cm]{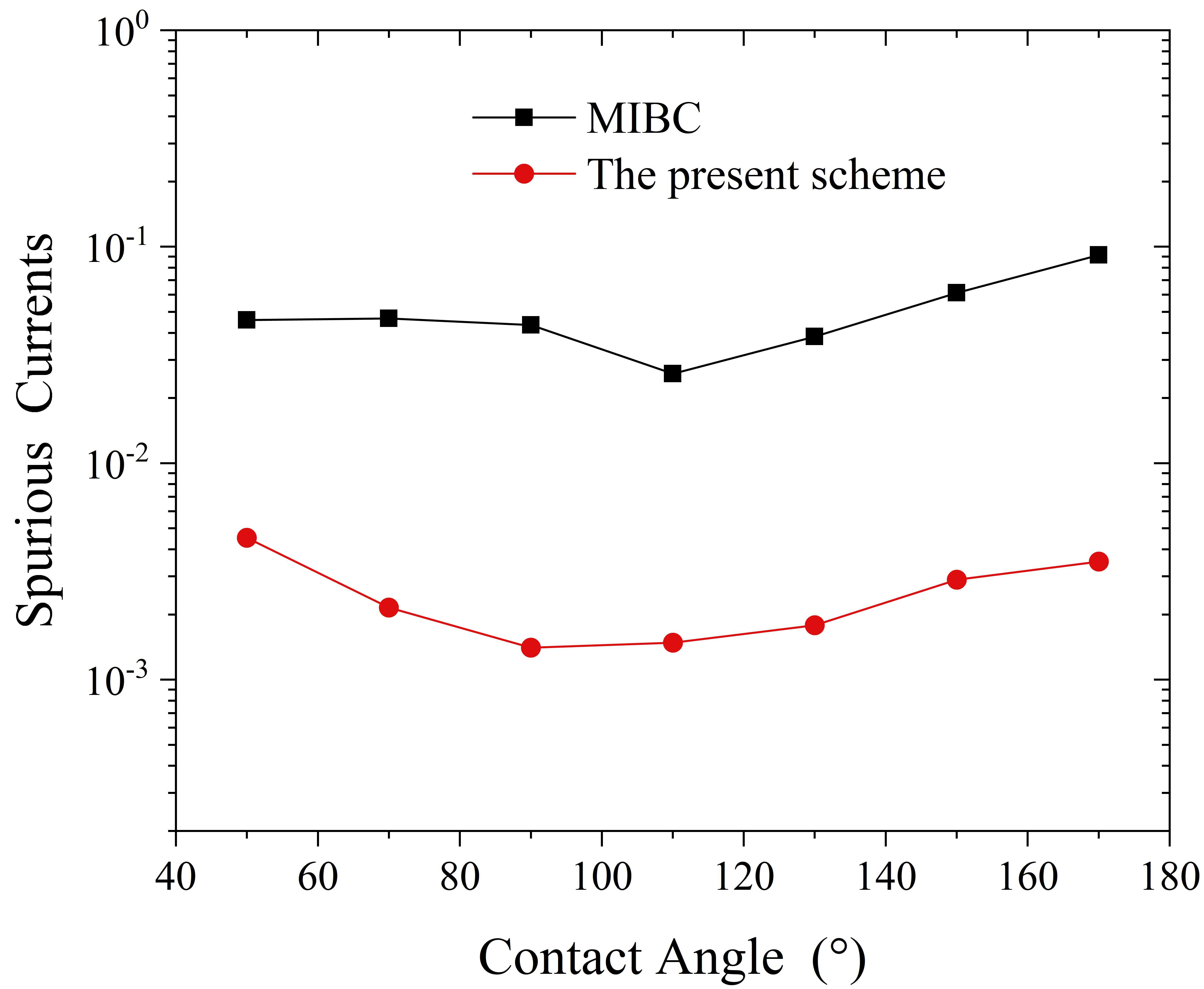}
	\caption{The spurious currents of a drop on the curved substrates with the wettability from hydrophilicity ($\theta = 50^\circ$) to super hydrophobicity ($\theta = 170^\circ$).}
	\label{FIG10}
\end{figure}

The spurious currents are further investigated by simulating a drop on the curved wetting substrates with a series of surface wettability. Since all kinds of $q$ values are involved, these simulations can reflect the overall effects of MIBC and the present scheme on the spurious currents. Fig. \ref{FIG10} presents that the spurious currents from the present scheme are one order magnitude lower than those from MIBC on all wetting substrates. These test cases validate the accuracy and robustness of the present scheme again.
\subsection{Drops falling on a curved wetting substrate}\label{sec4.5}
To further demonstrate the effectiveness of the present scheme in dynamic systems, we simulate the drop impact on a curved substrate, and the results are compared with the benchmarks from the experiments and simulations \cite{Bakshi2007,Fakhari2017}. The dimensionless parameters refer to the definitions in the literature \cite{Fakhari2017}. Bond number describes the ratio of gravitational to capillary forces and is defined by $Bo = \frac{{g\left( {{\rho _L} - {\rho _G}} \right){D^2}}}{\sigma }$, where $D$ is the diameter of the drop, $g$ is the gravity acceleration, $\sigma$ is the liquid/gas surface tension, ${\rho _G}$ and ${\rho _L}$ are the densities of the gas and liquid, respectively. The dimensionless time is related to gravity by $t_g^* = t\sqrt {g/D}$. In this section, we take $g = 980cm/{s^2}$ and $\sigma = 0.045$ for the simulations.
\begin{figure}[h]
	\centering
		\includegraphics[width=8.6cm]{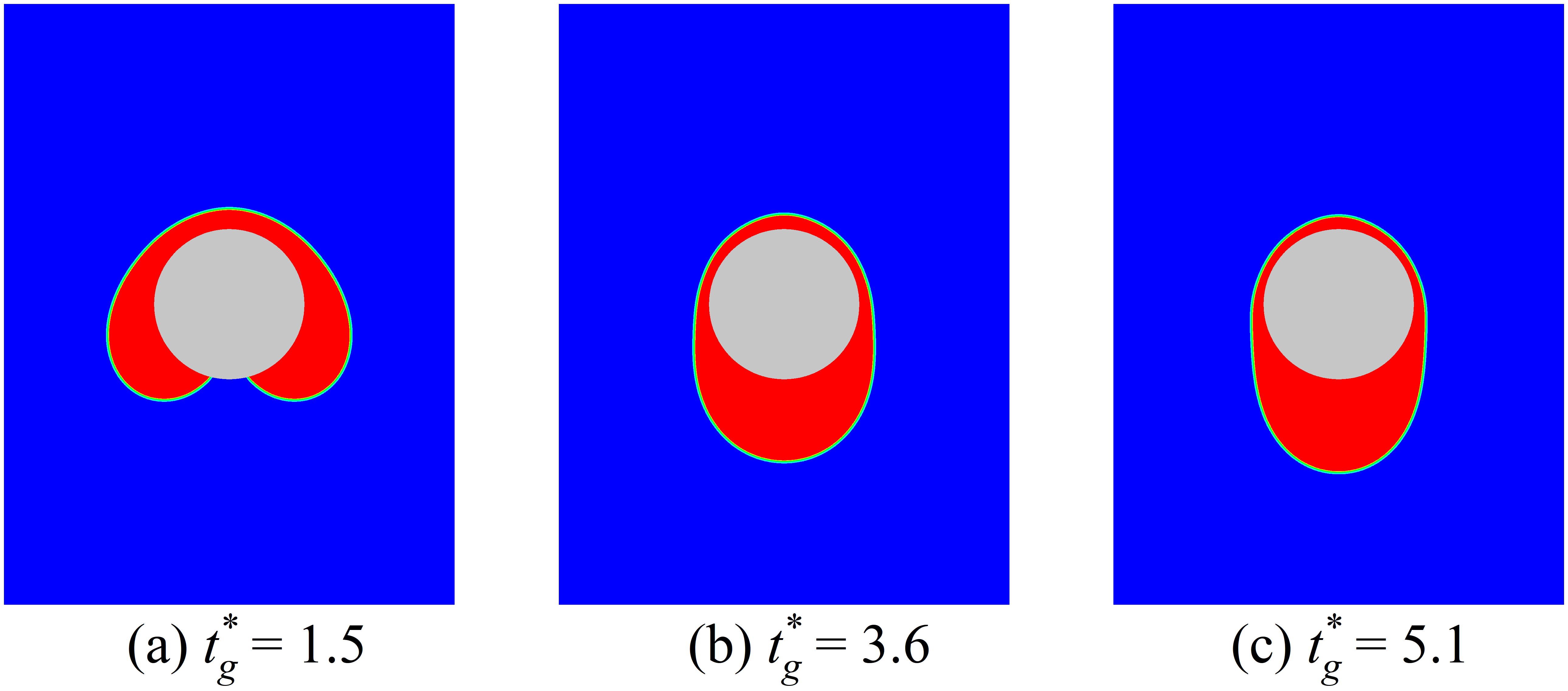}
	\caption{Dynamic evolutions of a drop impacting on a hydrophilic cylinder with $\theta $ = 60$^\circ$ and $Bo$ = 2.2.}
	\label{FIG11}
\end{figure}
\begin{figure}[h]
	\centering
		\includegraphics[width=8.6cm]{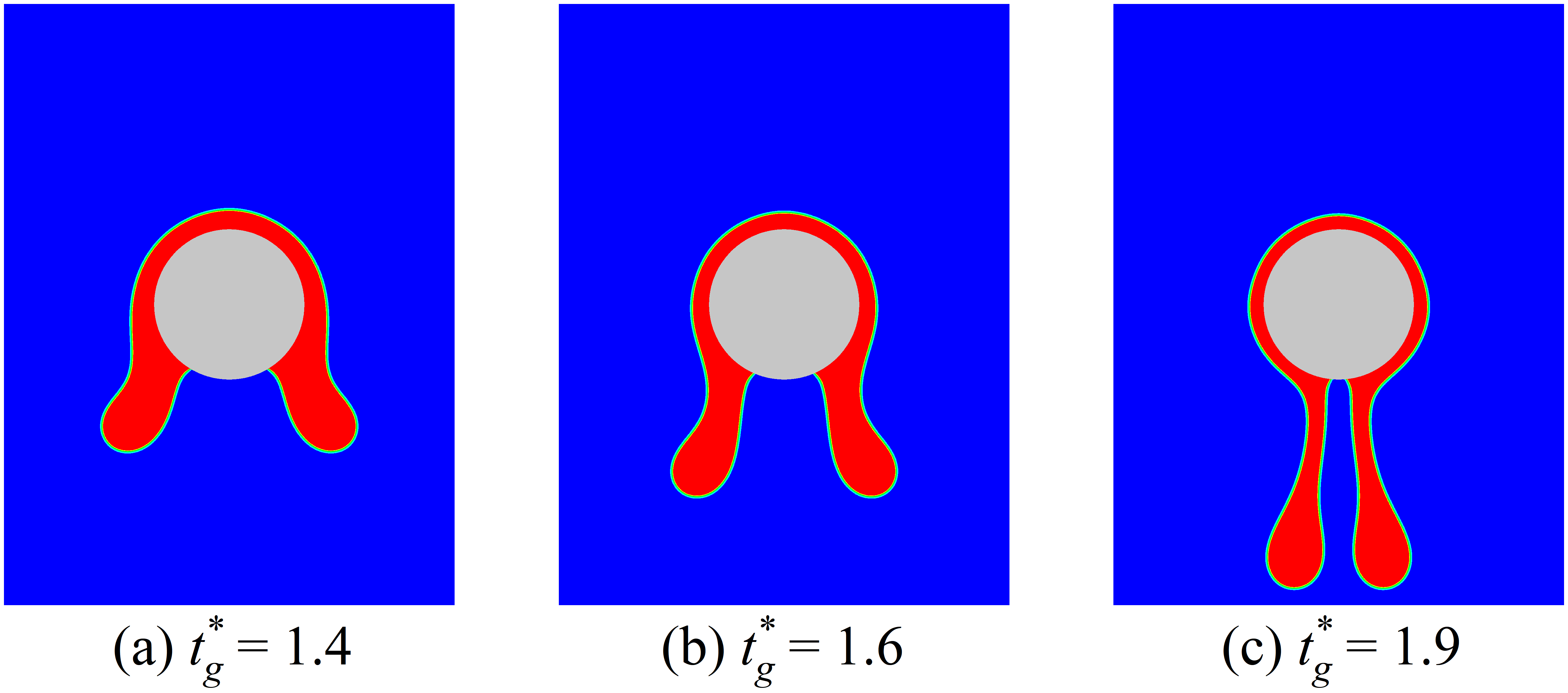}
	\caption{Dynamic evolutions of a drop impacting on a hydrophilic cylinder with $\theta $ = 60$^\circ$ and $Bo$ = 14.5.}
	\label{FIG12}
\end{figure}

Figs. \ref{FIG11} and \ref{FIG12} present a drop impacting on a hydrophilic substrate with $Bo$ = 2.2 and 14.5, respectively. In the evolutions, the drops tend to adhere to the substrate and engulf the perimeter of the hydrophilic cylinders due to the influence of gravity. Since the case in Fig. \ref{FIG12} has a larger $Bo$, it evolves faster, and the gravitational force is strong enough to prevail over the interfacial tension between the drop and the hydrophilic cylinder. This causes the drop to break up and detach from the hydrophilic substrate. 
\begin{figure}[h]
	\centering
		\includegraphics[width=8.6cm]{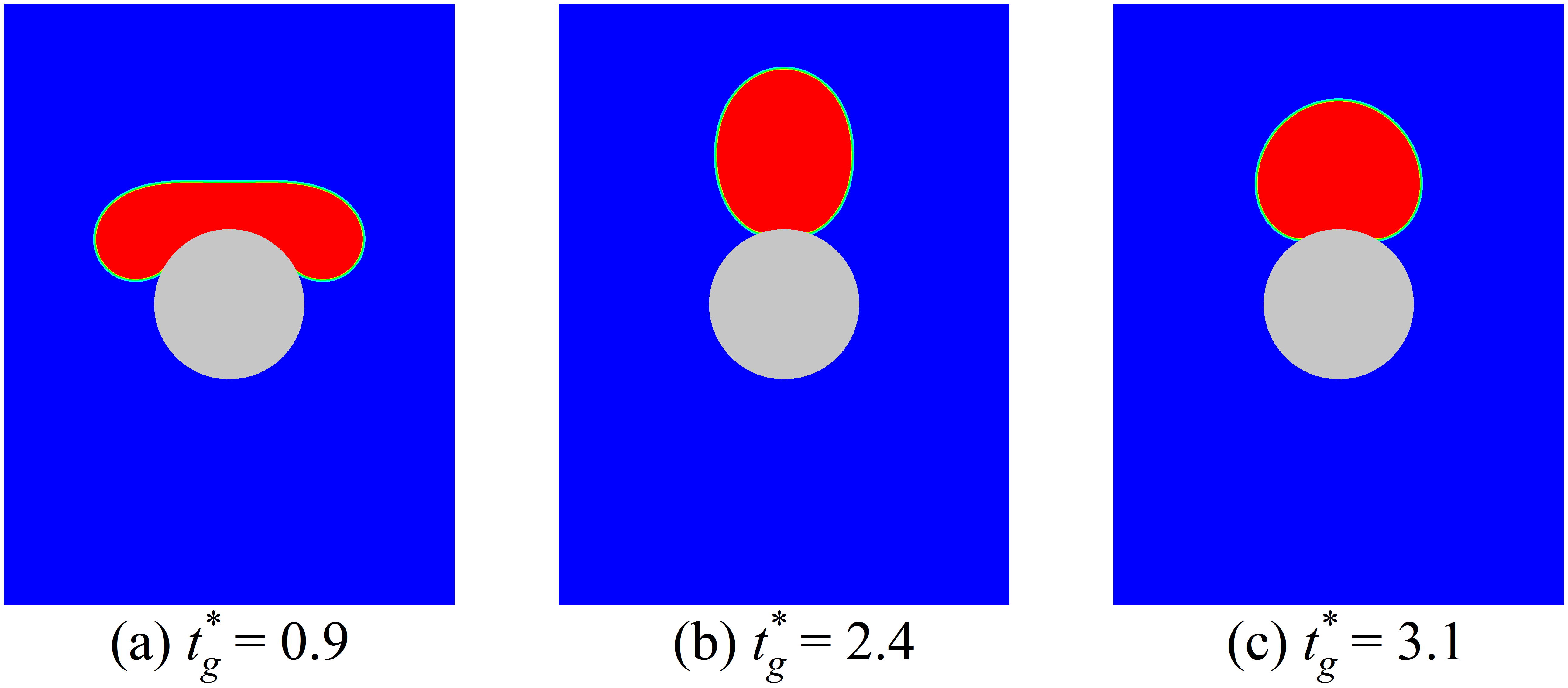}
	\caption{Dynamic evolutions of a drop impacting on a hydrophobic cylinder with $\theta = 150^\circ$ and $Bo$ = 2.2.}
	\label{FIG13}
\end{figure}
\begin{figure}[h]
	\centering
		\includegraphics[width=8.6cm]{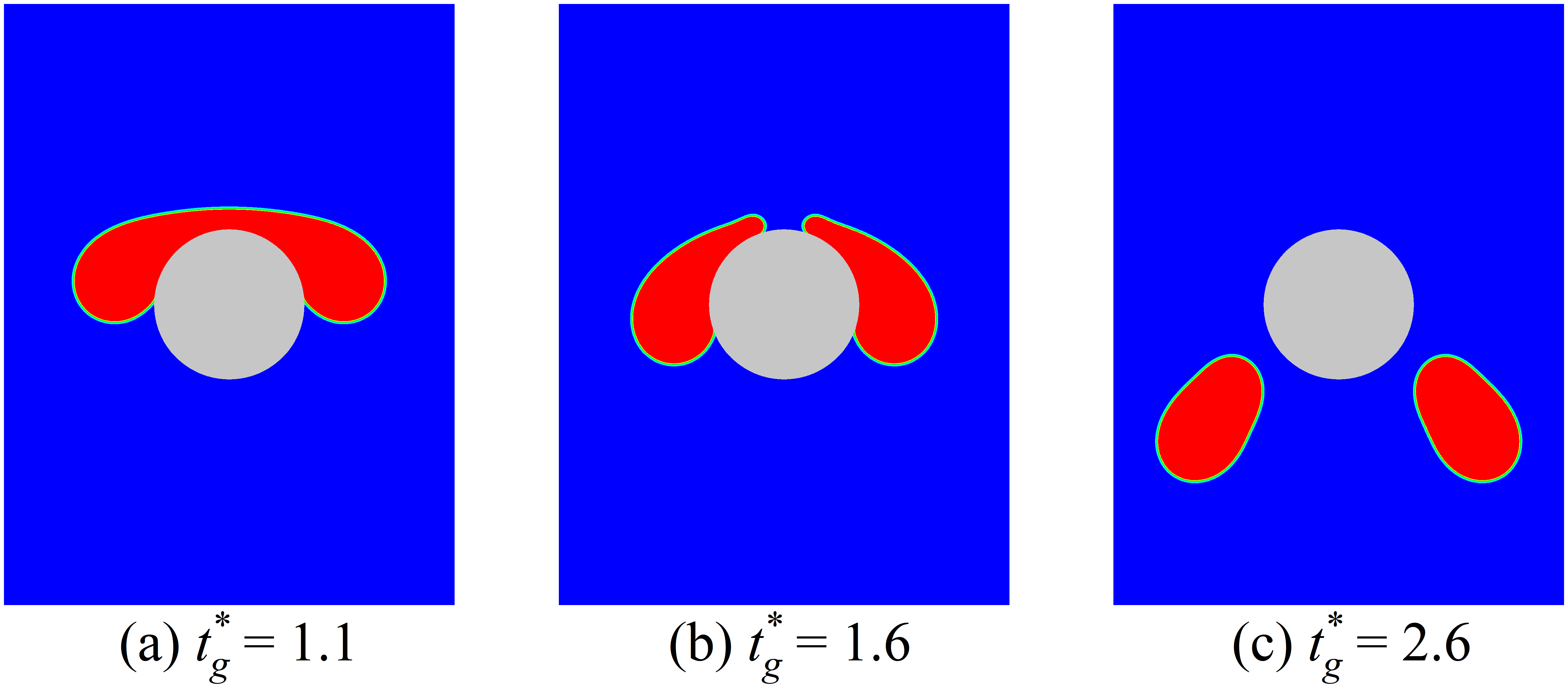}
	\caption{Dynamic evolutions of a drop impacting on a hydrophobic cylinder with $\theta = 150^\circ$ and $Bo$ = 4.0.}
	\label{FIG14}
\end{figure}
\begin{figure}[h]
	\centering
		\includegraphics[width=8.6cm]{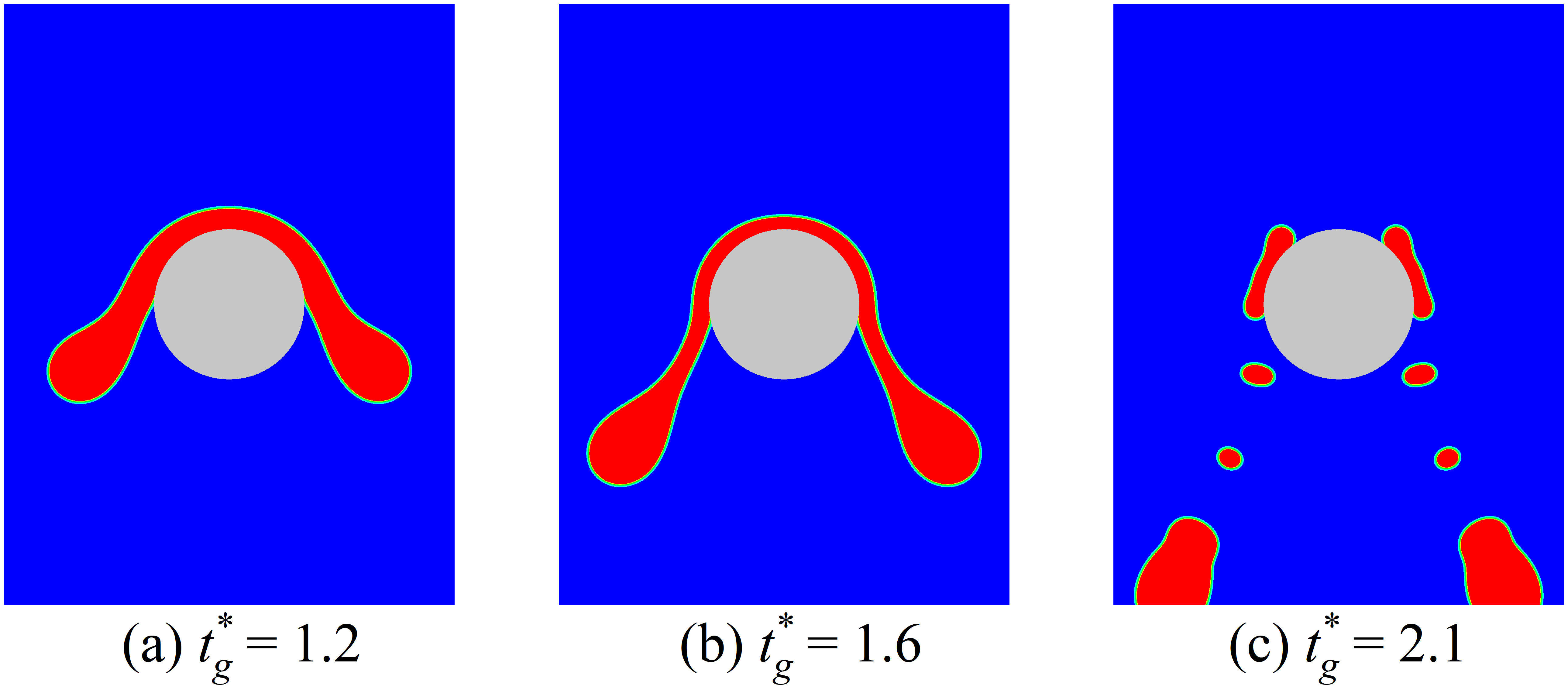}
	\caption{Dynamic evolutions of a drop impacting on a hydrophobic cylinder with $\theta = 150^\circ$ and $Bo$ = 14.5.}
	\label{FIG15}
\end{figure}

In order to compare the effect of the surface wettability, we increase the contact angle of the cylinder to $\theta = 150^\circ$, which corresponds to a hydrophobic substrate. The drop falls to the hydrophobic cylinder with three different Bond numbers, namely $Bo$ = 2.2, 4.0 and 14.5, and the simulating results are shown in Figs. \ref{FIG13} $\sim$ \ref{FIG15}, respectively. For the lowest $Bo$ = 2.2, Fig. \ref{FIG13} presents that the drop retracts after hitting the hydrophobic cylinder. This phenomenon is similar to the bouncing of drops on a flat surface \cite{Ji2021,Yarin2006}. Unlike Fig. \ref{FIG11}, in which the drop swallows the hydrophilic cylinder, the drop in Fig. \ref{FIG13} finally sits on the hydrophobic cylinder. With increasing $Bo$ to 4.0, the drop in Fig. \ref{FIG14} breaks upon contact with the upper pole of the substrate. Then, the drop is separated into two daughter drops and detached completely from the hydrophobic substrate. Further increasing $Bo$ to 14.5 results in an interesting disintegration. In the impact process shown in Fig. \ref{FIG15}, the drop is divided into several small drops of various sizes under the greater gravitational force. Similarly, the larger the $Bo$, the faster the evolution. 

\begin{figure}[h]
	\centering
		\includegraphics[width=8.2cm]{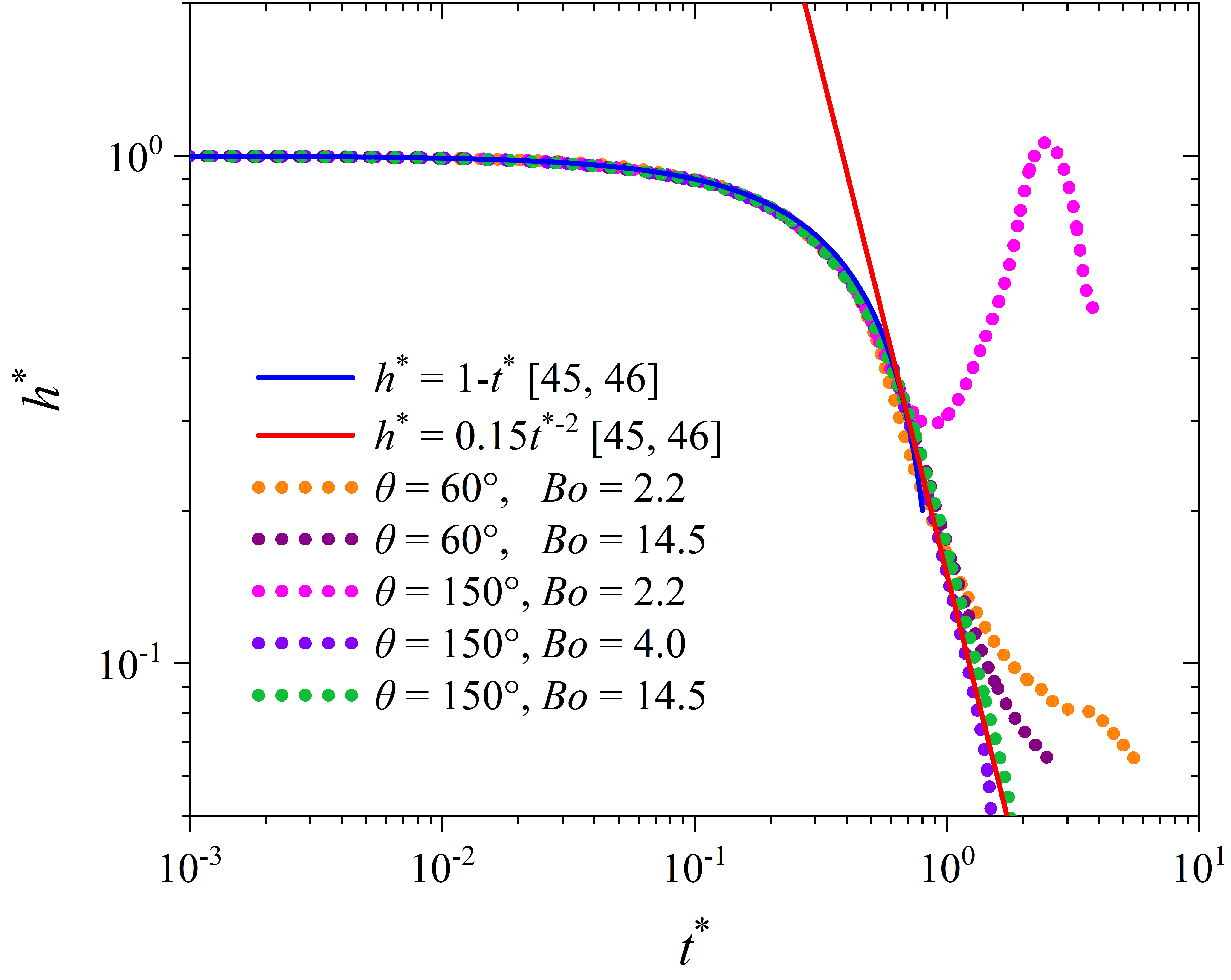}
	\caption{Temporal variation of the film thickness on top of the cylinder. The solid blue line and red line are the power-law curves from experimental data \cite{Bakshi2007,Fakhari2017}, and the dotted lines are the results of the present scheme simulation.}
	\label{FIG16}
\end{figure}

The film thickness ${h_f}$ on top of the cylinder can be measured and compared with the benchmarks from the experiments and simulations \cite{Bakshi2007,Fakhari2017}. The normalized film thickness is defined by ${h^*} = {h_f}/{h_i}$, where ${h_i}$ is the height of the drop at the instant of impact. 
A dimensionless time is defined as ${t^*} = \left( {t - {t_i}} \right){U_i}/D$, where ${t_i}$ is the instant of impact between the drop and substrate, and ${U_i}$ is the velocity at the moment of drop impact. The temporal variation of the film thickness on the top of the cylinder is plotted in Fig. \ref{FIG16}. According to the literature \cite{Bakshi2007}, the film thickness is expected to vary as ${h^*} = 1 - {t^*}$ at early times, and as ${h^*} = 0.15{t^{* - 2}}$ at intermediate times. The five simulations in Fig. \ref{FIG16} correspond to the dynamic evolutions in Fig. \ref{FIG11} $\sim$ \ref{FIG15}. It can be seen clearly that the simulation results by using the present scheme are in good agreement with the power-law fitted curves of the experimental data \cite{Bakshi2007}.

\section{Summary}\label{sec5}
The boundary treatment plays a significant role in the numerical simulations of complex fluid flows. In the lattice Boltzmann method, the effectiveness of the conventional curved boundary conditions in single-phase simulations has been confirmed by a lot of numerical results \cite{Mei1999,Bouzidi2001,Lallemand2003,Guo2002,Wen2014}. However, in multiphase environments, because the conventional curved schemes have not considered the nonideal effect of wetting boundaries and the interpolating error due to the remarkable density changes, they cause the dramatic mass leakage or increase, and usually crash the multiphase simulations. This paper proposes a multiphase curved boundary condition that incorporates the nonideal effect into the linear interpolation scheme and compensates for the interpolating error. The present scheme has been validated by a series of static and dynamic test cases with large density ratio. The numerical results show that it is accurate and guarantees mass conservation. It only requires a very small mass compensation, and the spurious current is suppressed to a very low level. These manifest that the present scheme is competent to depict the wetting boundary with complex geometry in the simulations of multiphase flow. We expect the multiphase curved boundary condition can promote the accuracy of the numerical simulations in the fields of capillary phenomena, moving contact line, soft wetting, etc. \cite{Sui2014,Liu2018,Andreotti2020}.

\begin{acknowledgments}
This work was supported by the National Natural Science Foundation of China (Grant Nos. 11862003, 81860635, and 12062005), the Key Project of Guangxi Natural Science Foundation (Grant No. 2017GXNSFDA198038), Guangxi “Bagui Scholar” Teams for Innovation and Research Project, and Guangxi Collaborative Innovation Center of Multisource Information Integration and Intelligent Processing. 
\end{acknowledgments}

\nocite{*}

%

\end{document}